\documentclass[
 reprint,
 amsmath,amssymb,
 dvipdfmx, floatfix,
 pra
]{revtex4-2}
\usepackage[dvipdfmx]{graphicx}
\usepackage{dcolumn}
\usepackage[T1]{fontenc}
\usepackage{bm}
\usepackage{booktabs}
\usepackage[normalem]{ulem}
\usepackage{color}
\usepackage{bbding}
\usepackage{sidecap}
\usepackage{multirow}
\usepackage{url}
\usepackage[dvipdfmx]{hyperref}
\usepackage{chngcntr}
\newcounter{lastequationbeforeappendix}

\begin{document}
\title[``Super-resolution'' optical tweezers]{Optimization-based hologram design for fine optical tweezers array and extension of super-resolution criteria}

\author{Keisuke Nishimura}
\email[Contact author:]{keisuke.nishimura@hpk.co.jp}
\affiliation{
  Hamamatsu Photonics K.K., Central Research Laboratory, Hirakuchi, Hamamatsu, Japan
}

\author{Hiroto Sakai}
\affiliation{
  Hamamatsu Photonics K.K., Central Research Laboratory, Hirakuchi, Hamamatsu, Japan
}

\author{Taro Ando}
\email[Contact author:]{taro@crl.hpk.co.jp}
\affiliation{
  Hamamatsu Photonics K.K., Central Research Laboratory, Hirakuchi, Hamamatsu, Japan
}

\author{Takafumi Tomita}
\affiliation{
Institute for Molecular Science, National Institutes of Natural Sciences, Okazaki, Japan
}

\author{Sylvain de L\'{e}s\'{e}leuc}
\affiliation{
Institute for Molecular Science, National Institutes of Natural Sciences, Okazaki, Japan
}

\date{\today}

\begin{abstract}
Aligning light spots into arbitrary shapes is a fundamental challenge in holography, leading to various applications across diverse fields in science and engineering. 
However, as the spot interval approaches the wavelength of light, interference effects among the spots become prominent, which complicates the generation of a distortion-free alignment. 
Herein, we introduce a hologram design method based on the optimization of a nonlinear cost function using a holographic phase pattern as the optimization parameter. 
We confirmed generation of a $5 \times 5$ multispot pattern, in which each spot is separated clearly by a distance of 0.952(1)~{\textmu}m, 
on the focal plane of a high-numerical-aperture (0.75) objective by observing the near-infrared (wavelength: 820~nm) holographic output light from a spatial light modulator device.
This result breaks through the spot spacing of a few micrometers that is a typical value achieved so far under similar experimental conditions. 
Furthermore, we propose redefinition of the Rayleigh diffraction limit by taking account of the separation of spot as well as the spot spacing.
The proposed method is expected to advance laser fabrication, scanning laser microscopy, and cold atom physics, among other fields.
\end{abstract}

\maketitle

\section{Introduction}\label{Intro}
Holography, which involves the control and utilization of both the amplitude and phase of light, has provided numerous benefits---ranging from academic to practical applications---to the modern society for more than seventy years \cite{Gabor1948, Gabor1951}. 
The technological advancement of optical devices has accelerated the progress of holography; notably, the spatial light modulator (SLM) allows for the direct control of light wavefronts through the optical properties of liquid-crystal molecules \cite{Lazarev2019}. 
Furthermore, the concept of a computer-generated hologram (CGH) facilitates the on-demand generation of desired light patterns by computing a phase pattern that reproduces the required output \cite{CGH2022}. 
Holographic methods that modulate the light phase using SLMs are widely applied because they enable dynamic pattern generation with high light utilization efficiency \cite{Lesem1969, Wyrowski1988}.

Among various applications of holography, generating light-spot arrays is one of the most popular examples for uses in laser fabrication \cite{sakakura2010, Sakakura2011}, scanning laser microscopy \cite{nikolenko2008}, and cold atom physics. 
Holographic optical tweezers are vital tools for aligning atoms in an array of arbitrary shapes \cite{nogrette2014, Browaeys2016, Lukin2021, IMS, IMS2}, in particular, to construct ``storage area'' in neutral-atom quantum computers \cite{Lukin2022, Bluvstein2024}. 
Generating short-spaced, clearly-resolved light-spot arrays with CGHs is expected to benefit those applications; however, this is a technical challenge addressing a general and fundamental issue in optics, namely, the realization of super-resolution \cite{PALM1, PALM2, STORM1, STORM2}.

CGHs for light-spot arrays are designed by a variety of methods, which are categorized into two main classes in terms of mathematical operations: self-consistent iterative methods and optimization-based methods.
The former ones are fundamentally based on the Gerchberg-Saxton algorithm (GSA) \cite{GS1972}, and in the field of hologram design, they are often referred to as iterative Fourier transform algorithms (IFTA) \cite{Wyrowski1988}. 
Popular CGH design methods such as weighted GSA (WGS) \cite{Ripoll2004review} and mixed-region amplitude-freedom (MRAF) \cite{Pasienski2008} methods are regarded as the iterative methods with additional tricks. 
The optimization-based methods were proposed as ones advantageous in convergence and reproducibility, and an individual method is identified according to the choice of a cost function, a quantity to be optimized, and an optimization algorithm, such as direct search, simulated annealing, and Newtonian method. An optimal-rotation-angle (ORA) method \cite{Bengtsson1994} is classified as a variation
of the optimization-based methods. 
All of the existing CGH design methods have own advantages but are, to a greater or lesser extent, ineffective for generating short-spaced light-spot arrays.

In this study, we propose a CGH design method that effectively generates a well-resolved fine multispot light array.  
The proposed method is based on the optimization of a cost function, with the holographic phase pattern serving as the optimization parameter.
The cost function is defined as the correlation between a given target pattern and the output light pattern, which results from the light propagation from the phase distribution in the CGH.
This choice of a nonlinear cost function brings drastic improvement of holographic generation of short-spaced light-spot arrays compared to a similar method \cite{harte2014}. 
By reproducing CGHs for a $5 \times 5$ multispot pattern with near-infrared light ($\lambda = 820$~nm), a spot spacing of $0.952(1)$~{\textmu}m was confirmed on the focal plane of a high-numerical-aperture ($\text{NA} = 0.75$) objective, a result which shows notable reduction from a few micrometers typically observed under similar experimental conditions so far.

Additionally, we establish a quantitative super-resolution criterion for evaluating fineness of a multispot light array. 
Well-known resolution criteria, i.e., Rayleigh, Sparrow, and Abbe criteria, are established upon the spacing and separation of two adjascent light spot, however, the criteria are reduced to a boundary value only of the spot spacing.
The proposed criterion is defined by the interrelationship between the spacing and separation, resultantly, to give a comprehensive one including those three criteria as well as importing the concept of modulation transfer function (MTF) \cite{MTF} by emphasizing the role of image contrast (or separation) in super-resolution.
Finally, the spot spacing obtained in this experiment does not satisfy the Rayleigh criterion, but analysis in conjunction with the spot separation showed that it satisfies our super-resolution criterion.

\section{Hologram design for fine spot arrays} \label{sec:cost_function}
\begin{figure}[tb]
    \includegraphics{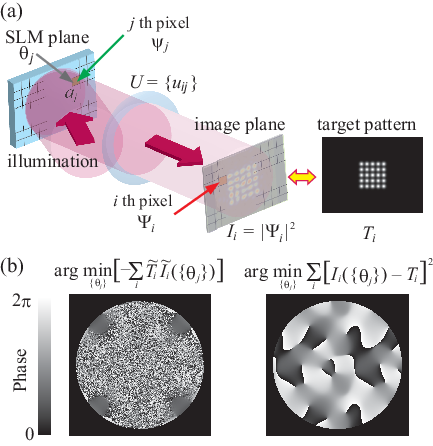}
    \makeatletter\long\def\@ifdim#1#2#3{#2}\makeatother
    \caption{
    (a)~Schematic of CGH design procedure. 
    On the SLM plane, user-defined phase $\theta_j$ is attached to incident light amplitude $a_j$ to make output light amplitude of SLM to be $\psi_j = a_j e^{i \theta_j}$.  
    Light propagation, expressed symbolically by an illustration of a lens, acts as a unitary transformation on $\psi_j$ to yield light amplitude on the image plane, $\Psi_i$. 
    Comparison of output pattern, $I_i = |\Psi_i|^2$, and target pattern, $T_i$, provides a guide to update $\theta_j$.
    (b)~CGHs for the 5$\times$5 square lattice pattern optimized by CGM using FOI (left) and RSS (right) cost function.
    }\label{fig:hologram_images}
    \end{figure}
Figure~\ref{fig:hologram_images}(a) illustrates the concept of CGH design using a reflection-type SLM. For simplicity, the two-dimensional position on the SLM plane is denoted by $j$, which can be regarded as a vector index. 
The SLM applies a phase shift $\theta_j$ to the incident light amplitude $a_j$ at the $j$th position. Consequently, the output light amplitude from the SLM is expressed as $\psi_j = a_j \exp(i \theta_j)$.

The position in the image plane can be specified by introducing another vector $i$ to approximately represent the light propagating from the SLM to the image plane which can generally be described by a unitary matrix $U = \{u_{ij}\}$, often referred to as the ``transfer matrix.''
This matrix is approximated by a Fourier transformation when the SLM and image planes are connected using imaging optics. 
Therefore, the light amplitude on the image plane $\Psi_i$ is related to $\psi_j$ as follows:
\begin{equation}
    \Psi_i = \sum_j u_{ij} \psi_j,  \label{eq:transfer}
\end{equation}
and the output pattern is $I_i = |\Psi_i|^2$. 
The CGH design is a problem of determining $\{ \theta_j \}$ for the desired target pattern $T_i$ under fixed illumination conditions corresponding to $\{ a_j \}$.

As described in Introduction, we attempt to extend optimization-based methods that offer higher computational stability and flexibility in designing CGHs with a drawback of requirement of more computational resources. 
IFTA \cite{Wyrowski1988}, the most popular example of the iterative methods, determines $\{\theta_j\}$ self-consistently by repeatedly performing light propagation calculations while imposing constraints on the light amplitudes on the SLM and image planes (Fig.~\ref{fig:hologram_images}{(a)}).
In contrast, the optimization-based methods determine the optimal CGH according to the degree of correspondence between $I_i(\{\theta_j\})$ and $T_i$.

The optimization-based CGH design method utilizing the analytical gradient of the cost function was first proposed by Harte \cite{harte2014}. 
In this approach, the set $\{ \theta_j \}$ is determined by optimizing the following cost function:
\begin{equation}
    f_\text{RSS}(\{ \theta_j \}) = \sum_i [I_i(\{ \theta_j \}) - T_i]^2, \label{eq:RSS}
\end{equation}
which involves minimizing the residual sum of squares (RSS) between $I_i$ and $T_i$. 
Note that $I_i$ and $T_i$ are normalized as $\sum_i I_i = \sum_i T_i = 1$, and $I_i$ is an explicit function of the phase distribution. 
As a merit of the optimization-based method, unique functionalities can be introduced into CGH design by selecting different cost functions. 
Bowman \textit{et al}.\cite{bowman2017} introduced a cost function based on a correlation between the target and output light amplitudes to achieve holographic complex-amplitude modulation. 
The final form of their cost function is a square of the light-amplitude correlation subtracted from unity, hence it can be regarded as a variation of RSS cost functions.
It is also noted that their aim is simultaneous modulation of amplitude and phase, not the generation of fine patterns. 
Here, we propose using simply the correlation of intensity, i.e., ``fidelity of intensity'' (FOI):
\begin{equation}
    f_\text{FOI}(\{ \theta_j \}) = - \sum_i \tilde{I_i}(\{ \theta_j \}) \tilde{T_i}, \label{eq:FOI}
\end{equation}
where $\tilde{ }$ denotes a quantity normalized with respect to its Euclidean norm.

In this study, we focus on unique and usable properties of Eq.~\eqref{eq:FOI} in comparison with Eq.~\eqref{eq:RSS} within the framework of optimization-based CGH design methods. 
Comparison of the proposed method with the self-consistent iterative methods is available via the previous study \cite{harte2014}, where properties of the optimization-based method with the choice of the RSS cost function [Eq.~\eqref{eq:RSS}] are presented in comparison with that of the IFTA method.
Additionally, adaptive modification, e.g., intensity feedback modification of the target pattern, is often introduced to each iteration step in self-consistent iterative methods for improving convergence and output image quality. 
Such kind of treatments are sometimes introduced in an ad-hoc manner and at a different level from the subject of this study; they are included in numerical optimization algorithms in the case of the optimizaion-based method. 
For this reason, the adaptive modifications are not addressed herein.

In the remaining part of this section, some remarks on numerical implementation of the proposed method are given.  
Among various numerical optimization algorithms, gradient methods become more advantageous in terms of computational complexity and convergence as the number of optimization parameters increases. 
In particular, the conjugate gradient method (CGM) is known as an efficient and stable one. 
While it is possible in principle to perform optimization by calculating gradients using numerical differentiation, we performed our calculations with analytical gradients because our design method becomes a large degree-of-freedom optimization problem involving a massive amount of computation. 
Fortunately, $\partial f_\text{FOI}/\partial \theta_j$ can be explicitly expressed as follows:
\begin{equation}
    \frac{\partial f_\text{FOI}}{\partial \theta_j}
    =\frac{2}{\mathcal{N}_I}
    \sum_i(-\tilde{T}_i + f_\text{FOI} \tilde{I}_i){\rm Im}[u_{ij}^* \Psi_j \psi_{i}^*], 
\end{equation}
where $\mathcal{N}_I (\equiv \sqrt{\sum_i I_i^2})$ represents the Euclidean norm of $\{ I_i \}$.  
This expression enables efficient coding that completes large-scale optimization in a reasonable time, starting with an arbitrary random pattern.

Numerical optimization yielded holographic phase patterns as two-dimensional arrays of double-precision floating-point numbers within the interval $[0, 2\pi)$. 
As the SLM device is designed to handle 8-bit signals, the phase values were discretized into 256-step (8-bit) quantities for practical application. 
In this study, ``CGHs'' refers to the discretized versions.
In a practical CGH design, the SLM plane is typically represented as an $N \times N$ square grid, where $N$ is a positive integer. The active area of the SLM is embedded within this grid by setting $a_j = 0$ outside the active area. 
The image plane also becomes an $N \times N$ grid because the action of $U$ preserves the grid size. Suppose hologram reproduction is performed using light of wavelength \(\lambda\) with imaging optics of effective focal length $f$. In that case, the actual size of the target can be estimated using the relation $\Delta x^\prime = \lambda f /(N \Delta x)$, which describes the relationship between the grid spacings on the SLM and image planes, $\Delta x$ and $\Delta x^\prime$, respectively.

\section{Experiments}\label{sec:experiments}
\begin{figure}[htb]
  \includegraphics{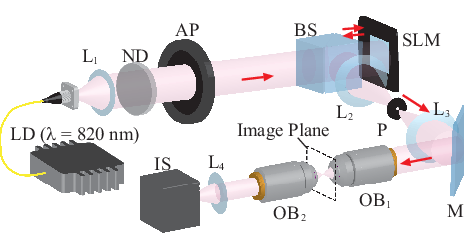}
  \makeatletter\long\def\@ifdim#1#2#3{#2}\makeatother
\caption{
Schematic of experimental setup for hologram reproduction.  
LD: laser diode, $\text{L}_1$-$\text{L}_4$: achromatic lens, ND: neutral density filter, AP: aperture, BS: beam splitter, P: pinhole, M: mirror, $\text{OB}_1$-$\text{OB}_2$: objective lens, IS: CMOS image sensor.
} \label{fig:system}
\end{figure}
Experiments were conducted using the optical system depicted in Fig.~\ref{fig:system}. 
An 820~nm laser diode with a fiber port (LP820-SF80, Thorlabs) served as the light source.
Output light from fiber port was collimated using an achromatic lens and was shaped through an aperture. 
The collimated light was directed onto a reflection-type liquid-crystal-on-silicon SLM (X13138-01 modified, Hamamatsu). 
Phase holograms [Fig.~\ref{fig:hologram_images}(b)] were displayed on the SLM surface that is filled up with $1272 \times 1024$ pixels (pixel size: 12.5~{\textmu}m $\times$ 12.5~{\textmu}m).
A blazed phase grating pattern was superimposed to demarcate the non-modulated (or, zeroth-order) light components from SLM outputs \cite{Matsumoto2008}.
Since the blazed phase pattern was displayed inside a 360~px-radius circle around the center of the SLM surface, a radius of the effective aperture of the holographic outputs became 4.5~mm. Noting that the Gaussian radius of the illumination profile was estimated as approximately 20~mm on the SLM surface, the relative standard deviation of the illumination can be estimated as less than 0.030 over the effective aperture.
Thus, the incident light profile can be regarded as a pseudo top-hat.

For generating holographic light-spot arrays, the SLM outputs were separated by a beam splitter and transferred through a 4$f$ telescope---0.8X-magnification optical system consisting of two achromatic lenses with a focal length of 125~mm and 100~mm---to the entrance pupil of an objective lens [MRH00401, Nikon: NA = 0.75 (air)]. 
If the illuminated area on the SLM plane (i.e., the SLM pixels labeled by $\{ \forall j | a_j \neq 0\}$) is not incident on the entrance pupil of the objective lens, it reduces the effective NA and fails to generate a short spot spacing, thereby limiting the performance of the objective. 
Therefore, a suitable choice of the illumination area and/or magnification of the transfer optical systems was required. 
Focal images were observed using a 60X-magnification optical system, comprising the same objective lens used for focusing, an achromatic lens with a focal length of 300~mm, and an image sensor (Lumenera, LU135M-IO-WOIR).

We designed CGHs by applying the CGM to minimize the FOI cost function for given target patterns. The number of CGM iterations was set at 1000, which is typically sufficient for the relative decrease in FOI cost per iteration step to be less than $10^{-8}$. The minimization-based hologram design problem does not necessarily yield unique solutions. Therefore, the designed CGHs depend on the initial patterns selected, which are prepared by assigning a pseudorandom number (in the range $[0, 2\pi))$ to each lattice point on the grid in the SLM plane.

Here, we assume that the hologram and the target planes are connected using Fourier optics.
The input and target planes are represented as spatial grids of the same size in the numerical Fourier transform, with the target plane comprising 800$\times$800 points.
To increase the effective resolution of the target image by a factor of ten, the method described in \cite{kim2019gerchberg} was applied to the SLM plane.
Specifically, the hologram region is embedded into the center of an 8000$\times$8000 grid, where the light amplitude is set to zero outside the hologram.
Moreover, we assume that the illumination is a top-hat laser beam, indicating that $a_j$ in Fig.~\ref{fig:hologram_images}(a) presumes a nonzero constant value within the central circular region of a 360-pixel radius.

Ultimately, the hologram design becomes a 640,000-dimensional numerical optimization problem. 
Most computing resources are consumed by repeated evaluations of the cost function and its derivatives, which involve the propagation calculations of the two-dimensional light amplitude distribution expressed by 8000$\times$8000 components. 
To efficiently perform such large-scale propagation computations, we applied a double-precision complex FFT algorithm running on general-purpose graphics processing units, achieving reasonable computational times (typically 10--20 minutes).

Figure~\ref{fig:hologram_images}(b) illustrates examples of the designed CGHs obtained using the FOI and RSS methods, respectively. 
The CGH designed with the FOI method appears to exhibit an irregular and rough phase front to raise a question whether it is merely a mathematical object without physical meaning.
Nevertheless, as will be comfirmed later, the FOI method is superior to the RSS method in terms of both uniformity and separability.

\begin{figure}[htb]
  \includegraphics[width=\columnwidth]{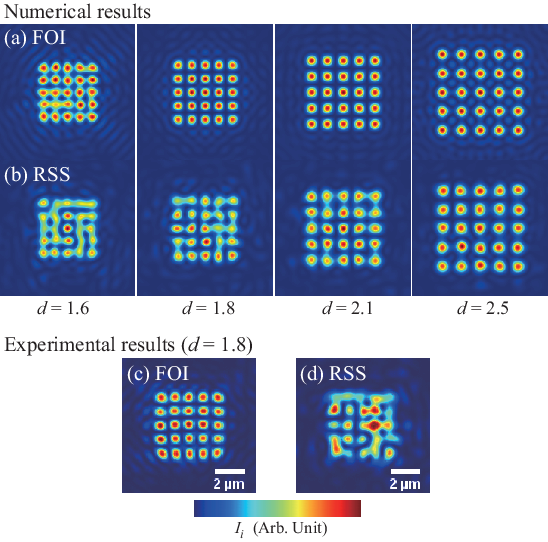}
  \makeatletter\long\def\@ifdim#1#2#3{#2}\makeatother
  \caption{
  (a), (b): Numerically reproduced images of square lattice target patterns with different spot spacings.
  The upper row was calculated from FOI-designed CGHs, and the lower row from RSS-designed ones.
  The spot spacings are 1.6, 1.8, 2.1, and 2.5 pixels from left to right.
  (c), (d): Experimental results of images reproduced from FOI- and RSS-designed CGHs, respectively.
  The spot spacing was set to 1.8~pixels in a common target pattern,
  corresponding to an actual length of $0.952(1)$~{\textmu}m on the image plane.
} \label{fig:rep_images}
\end{figure}

\begin{table*}[htb]
  \caption{
  Properties of reproduced spot arrays. 
  The spatial uniformity is assessed by the relative standard deviation (SD), which is the ratio of the SD to the mean value, in the powers of the $5 \times 5$ light spots that form the spot array. 
  The light utilization efficiency is calculated as the ratio of the output light power (measured by a power meter through a 1-mm-diameter pinhole placed on the image plane) to the illumination power measured on the SLM plane. 
  The light utilization efficiency is simply stated as the sum of $I_i$ over the area displayed in Fig.~\ref{fig:rep_images} for numerically reproduced images. 
  The value and error of a physical quantity indicate the mean value and standard error (SE), respectively, in a series of quantities evaluated numerically or experimentally from $20$ CGHs, each designed by starting from a different initial random pattern to generate a common target.
  For numerical calculations with aberrations, see the main texts in Sec.~\ref{sec:Results_uniformity}.
  }
  \label{tab:var_eff}
  \centering
  \begin{tabular*}{\textwidth}{@{\extracolsep\fill}lcccc}
  \toprule
   & \multicolumn{2}{@{}c@{}}{Spot uniformity} & \multicolumn{2}{@{}c@{}}{Utilization efficiency} \\
  \cmidrule{2-3}\cmidrule{4-5}%
   & FOI & RSS & FOI & RSS \\
  \midrule
  Numerical & $1.21(4) \times 10^{-2}$ & $5.6(10) \times 10^{-2}$ &  0.120(3)  &  0.928(4)    \\ 
  Numerical (with aberrations) & $5.2(2) \times 10^{-2}$ & $1.30(16) \times 10^{-1}$ &  0.119(3)  &  0.928(4)    \\ 
  Experimental & $9.70(15) \times 10^{-2}$ & $2.7(3) \times 10^{-1}$  & 0.127(2)  &  0.64(3)  \\ 
  \botrule
  \end{tabular*}
\end{table*}

\section{Results}\label{sec:results}

\subsection{Hologram reproduction and analyses}
To investigate the properties of the proposed method, the reproduced images of RSS- and FOI-designed CGHs were compared for target patterns of $5 \times 5$ square spot arrays. 
Figures~\ref{fig:rep_images}(a) and (b) respectively show numerically reproduced images of the RSS- and FOI-designed CGHs for target patterns with spot spacings of 1.6, 1.8, 2.1, and 2.5~pixels. 
When the spot spacing of the target pattern becomes less than 2.1~pixels, the spots in the numerically reproduced images begin to merge for the RSS-designed CGHs, whereas the results of FOI-designed ones maintain clear separation. 
However, even in the numerical reproduction, the spot separation breaks down in the reproduced images of the FOI-designed CGHs for target spot spacing of less than 1.8~pixels.

The output images reproduced from CGHs were observed using the experimental setup illustrated in Fig.~\ref{fig:system}.
Spot spacing of target patterns are set to 1.8~pixels, which corresponds to 0.9225~{\textmu}m in the image plane, for observing a distinct difference between results of the FOI and RSS methods.
We generated 20 holograms by applying both RSS and FOI methods starting with different random patterns as initial conditions in optimization calculation. 
Figures~\ref{fig:rep_images}(c) and (d) display examples of reproduced images of the FOI- and RSS-designed holograms, respectively.
We also tried to reproduce these images for target spot-interval of less than 1.8~pixels experimentally, but are noticed that it would be difficult to obtain well-resolved multispot patterns.

The spacing and positions of the spots were determined by applying fitting analysis to the output patterns on the image plane.
We adopt the following model function for an incoherent superposition of Gaussian spots with a common spot radius, $w_0$, on uniform background $B$:
\begin{equation} \label{eq:fitting}
    I_\text{fit} = \sum_k A_k \exp\left[-\frac{(x - x_k)^2+(y - y_k)^2}{w_0^2/2}\right] + B,
\end{equation}
where $k$ denotes an index to distinguish the spots, and the peak ``height'' and position of the $k$-th spot is expressed as $A_k$ and $(x_k, y_k)$.

The tilt angle of an image, $\theta$, is included in the practical fitting calculations.
Consequently, the fitting model becomes a (three times the number of spots plus three)-dimensional problem, rendering the analysis as computationally demanding.
Therefore, the quantitative analyses were restricted to a $5 \times 5$ square spot alignment with spacing $d$. 
After positioning a peak at $(X_0, Y_0)$, the other $(x_k, y_k)$ positions are automatically determined as $(X_0 + m d, Y_0 + n d)$ for pairs of suitable integers $m$ and $n$, depending on the geometrical configuration of the target spot array.

The number of trials in the analyses of RSS results was reduced because one of the CGHs obtained using the RSS method was excluded from the experiments and analyses due to the failure of convergence in the hologram design.
Although the RSS method successfully yields CGHs, the reproduced images of the RSS-designed holograms exhibit significant distortions, as depicted in Figs.~\ref{fig:rep_images}(b) and \ref{fig:rep_images}(d).
In such cases, the aforementioned analysis often fails to determine the peak positions in the reproduced images.
Therefore, we adopted the results of an FOI-designed hologram that was obtained using the same initial random pattern.

The mean spot spacing is 0.952(1)~{\textmu}m for Fig.~\ref{fig:rep_images}(c), which cannot be attained by any other existing hologram design methods under similar optical conditions. 
Actually, reproduced patterns on the focal plane was observed by super-resolution optical pattern measurement \cite{tomitaICAP2022, tomita2024} with the consequence that the spot spacing was estimated as less than 1~{\textmu}m using 852 nm trapping beam with a high-NA objective (NA=0.75 in design).
Although adaptive adjustment of CGHs is frequently incorporated into CGH designs to enhance the image quality of holographic outputs \cite{Matsumoto2012, Poland2014}, the current results were obtained without any post-modification.

Preliminary experiments on fluorescence image observation also confirmed trapping of atoms on the focal plane.
These results will be reported elsewhere in the future.

\subsection{Spot uniformity and light utilization efficiency}\label{sec:Results_uniformity}
Considering the scope of this study, the evaluation of the reproduced images should be based on the characteristic properties of light-spot arrays. 
One might assume that a value of FOI itself serves as a measure of the quality of the reproduced images, but the FOI reflects geometrical information of light-spot arrays only in an indirect manner.
Therefore, we investigated the properties of the reproduced images in terms of the uniformity of the light-spot array and light utilization efficiency.

The spot uniformity and utilization efficiency of the images reproduced using FOI- and RSS-designed CGHs are presented in Table~\ref{tab:var_eff}.
Spot uniformity is derived as the ratio of the standard deviation to the mean light powers of the twenty-five spots. 
Here, we use fitting parameters $\{x_k, y_k, w_0\}$ in Eq.~\eqref{eq:fitting} for deriving the power of the $k$th light spot in an observed or numerical spot-array image. 
Once the position of the $k$th spot, $(x_k, y_k)$, and the Gaussian radius of each spot, $w_0$, are determined, the power of a light spot is given as a sum of pixel values within an interval of $[x_k - w_0/2, x_k + w_0/2] \times [y_k - w_0/2, y_k + w_0/2]$ in the observed image. 
The results reveal that the FOI method exhibits small variations in spot power but demonstrates low light utilization efficiency, despite employing phase-only modulation. In contrast, the RSS results display the opposite behavior. 
Spot uniformity was suppressed in both numerical and experimental results compared to FOI, while light utilization efficiency shows higher values in both results. 
Thus, the FOI method can generate high-quality spot arrays in terms of uniformity and stability, although at the expense of efficiency.

We also examined the effects of residual aberrations on the spot uniformity and light utilization efficiency. 
Recent literatures by one of the authors reported that even high-quality objective lenses cannot avoid residual aberrations of less than 0.02~$\lambda$ in root mean square \cite{Otsu2024, Ando2024}, which is far below detection limits of any existing methods.
We calculated and analyzed the reproduced images by adding a typical aberration pattern observed for a plan-fluorite--type objective
(Objective A in Ref.~\onlinecite{Ando2024}) to 20 FOI- and RSS-designed holograms.
The results are shown in the middle row of Table~\ref{tab:var_eff}, suggesting that aberrations can certainly be a factor degrading the spot uniformity.
However, it is noted that the objective lens and optical train in this study are strictly different from those in Refs.~\onlinecite{Otsu2024} and \onlinecite{Ando2024}, and that these results are for informational purposes.

\begin{figure}[tb]
  \includegraphics{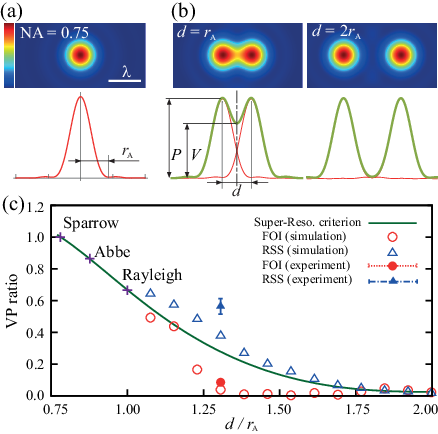}
\makeatletter\long\def\@ifdim#1#2#3{#2}\makeatother
\caption{
(a)~Focal pattern (top) and profile (bottom) of Airy disk [NA = 0.75(air)].
Red line in the bottom plot shows cross-sectional profile corresponding to the focal pattern, where $r_{\text A}$ refers to the Airy-disk radius. 
(b)~Focal patterns and profiles of incoherent sums of two Airy disks aligned at distances of $d = r_{\text A}$ (left) and $2r_{\text A}$ (right). 
In the bottom plots, thick green lines display the profiles of the incoherent sums, whereas thin red lines indicate those of individual Airy disks. 
Definitions of peak (P) and valley (V) are illustrated as well.  
(c)~The relationship between the VP ratio and spot spacing for incoherent sums of adjacent Airy disks is depicted (solid green line), 
where the traditional Sparrow, Abbe, and Rayleigh criteria are plotted as purple crosses. 
Open symbols represent VP ratios for numerically reproduced images from FOI- (red open circles) and RSS-designed (blue open triangles) CGHs, while closed symbols represent those for the corresponding observed patterns, i.e., Figs.~\ref{fig:rep_images}(c) (red closed circle) and (d) (blue closed triangle), respectively. 
Error bars (SEs) are determined similarly to those in Table~\ref{tab:var_eff} but are smaller than the size of the plot symbols, except for the experimental result of an RSS-designed CGH.}\label{fig:inversePV} 
\end{figure}

\subsection{Relation between spacing and separation of spot array}
FOI-designed holograms can generate high-quality spot arrays with extremely narrow spacing. 
To evaluate the behavior of spot arrays within the diffraction-limit regions, the focal patterns calculated under a high-NA [= 0.75 (air)] condition are displayed in Fig.~\ref{fig:inversePV}. 
Note that all numerical results in Fig.~\ref{fig:inversePV} were calculated using vector Debye formulas \cite{Richards1959, Boivin1965, Youngworth2000, PhysRevLett.99.073901, PhysRevA.79.033830}, as the focal patterns of coherent light are affected by the vector nature of the light amplitude under NA conditions greater than 0.70. 
The Airy disk pattern calculated assuming circularly polarized light is illustrated in Fig.~\ref{fig:inversePV}(a).

The Rayleigh-limit conditions portrayed on the left in Fig.~\ref{fig:inversePV}(b) present the adjacent light spots (Airy disks) as inadequately separated, while well-separated light spots, as displayed on the right of Fig.~\ref{fig:inversePV}(b), are desirable for maintaining a high aspect ratio in applications such as laser processing and suppression of background noise in scanning laser microscopy.
Therefore, a factor other than spot spacing is required for a proper evaluation of spot-array patterns while considering their applications.

The valley-to-peak (VP) ratio of a spot array along with the inverse of the peak-to-valley ratio was introduced as a metric for spot separation.
A smaller VP ratio indicates a clear separation of adjacent light spots. The VP ratio is advantageous because of its availability and robustness in derivation under unfavorable conditions compared to other metrics, such as the size of each light spot.
To derive the VP ratio of the reproduced spot-array image, the peak positions were determined by fitting the reproduced images. 
The peak value refers to the mean of $I_i$ at the peak positions of all the bright spots, whereas the valley value refers to the mean of $I_i$ at all the midpoints between the adjacent peaks. 
Thus, the VP ratio was evaluated as the ratio of the valley value to the peak value.

The blue (red) open symbols displayed in Fig.~\ref{fig:inversePV}(c) represent the VP ratios of numerically reproduced images from hologram patterns designed by the FOI (RSS) method, whereas the blue (red) closed symbols indicate those derived from the experimental images. 
In terms of the VP ratio, the proposed FOI method outperforms the conventional RSS method in the range of $(1 < d/r_\text{A} < 2)$.

\section{Discussion}\label{sec:discussion}

\subsection{Extension of super-resolution criterion}\label{sec:criterion}
Now we attempt to extend the known super-resolution criteria to a definite, quantitative, and comprehensive one.
In considering the relationship between physical quantities affecting spot separation, the traditional Rayleigh diffraction limit is established for an incoherent superposition of two Airy disks.
Thus, this approach should be logically followed.
We propose to determine whether a spot array exhibits ``super-resolution'' by examining the relationship between the spacing $d$ and the VP ratio of two incoherent Airy disks, as illustrated by the solid green line in Fig.~\ref{fig:inversePV}(c). 
This criterion quantitatively delineates a limit that cannot be surpassed using incoherent light and is beneficial for discussing the separability of multispot patterns using coherent light.

The area below the green line can be regarded as a super-resolution region, where the separation of adjacent spots is superior to the criterion for fixed spot spacing, or where the spot spacing is smaller than the criterion for a fixed VP ratio. 
According to the extended super-resolution criterion, the numerical VP values in Fig.~\ref{fig:inversePV}(c) suggest the potential of FOI-designed CGHs to generate super-resolution light-spot arrays in the regime $d/r_\text{A} < 1.8$. 
In particular, superior spot separation is anticipated for $1.3 < d/r_\text{A} < 1.7$. 
In this interval, the $d/r_\text{A}$-VP plot for the results of the RSS-designed CGHs lies around the extended criterion but not in the super-resolution region.
On the other hand, in the region of $d/r_\text{A} > 1.8$, the VP ratios of the FOI and RSS method do not differ significantly, and this trend continues for $d/r_\text{A} > 2.0$, outside the range shown in Fig.~~\ref{fig:inversePV}(c).
So, the RSS method will be preferable in terms of the light utilization efficiency under the condition of larger spot spacing.

The experimental VP ratios at $d/r_\text{A} = 1.4$ in Fig.~\ref{fig:inversePV}(c) exhibit moderately larger values than the numerical ones, suggesting the influence of extrinsic factors in the experiments, such as residual optical aberrations. 
Nevertheless, we emphasize that the results of the FOI method remain within the super-resolution regime.
Refer to Supplemental Material \cite{Supple} for further details.

So far, the discussion has been based on the Rayleigh criterion, but there are other super-resolution criteria, such as the Sparrow and Abbe criteria.
Difference among those three criteria is reduced to the definition of separability bewteen two adjascent light spot; i.e., they differ only in the choice of the limit spacing at which the two Airy disks can be regarded as resolved. 
Our super-resolution criterion can be said to encompass all three of these, since all three criteria correspond to a single point (purple cross) on the solid green line in Fig.~\ref{fig:inversePV}(c).

For example, the strictest Sparrow criterion, below which the valley between the two superimposed Airy disks disappears, can be expressed as a point corresponding to VP ratio of unity. 
At spot spacings closer than this, the valley value exceeds the peak, making the VP ratio as a physically meaningless measure.
Thus, the Sparrow criterion also indicates the minimum spot spacing at which our super-resolution criterion is applicable. 

Since this paper focuses on establishing and demonstrating the fundamentals of super-resolution criterion, the target pattern has been limited to square lattices. 
Therefore, quantitative discussion of non-square lattice patterns is not included here, but comments on applicability of our method to these patterns is provided in Appendix~\ref{sec:NonSquareLattice}.

\subsection{Stability of CGHs against external disturbance}\label{sec:stability}
A sequence of pseudorandom numbers is completely determined by selecting an integer, known as the ``random seed.''
Different seeds can be selected to generate various hologram patterns for the same target patterns. 
The images reproduced from these different hologram patterns can be treated as independent statistical trials to evaluate the statistical errors of the physical quantities obtained through analyses.

In this study, the error of a physical quantity, defined as SE in a series of quantities, provides a benchmark for the stability of the reproduced images. 
It includes the effects of the non-uniqueness of the hologram design problem, the convergence criterion set for optimization in CGH design, the handling of phase values in CGHs (rounded to discrete values), and the numerical/experimental accuracy for hologram reproduction.
The accuracy of the numerical holographic reproduction was equivalent to that of the FFT algorithm and is expected to be negligibly small (less than $5 \times 10^{-5}$ relative to the root-mean-square error, even for single-precision FFT, as noted in \url{https://pypi.org/project/pyvkfft/}).
This fact justifies neglecting the errors of numerically derived quantities.
Therefore, the instabilities in the hologram design and optical reproduction stages, namely, the intrinsic and extrinsic instabilities, can be distinguished by comparing the numerical and experimental results presented in Table~\ref{tab:var_eff}.

As the first point of issue in Table~\ref{tab:var_eff}, errors in the numerical light utilization efficiencies were similar for both the FOI and RSS methods.
This observation suggests that the accuracy of the CGM procedure, the only common factor between the FOI and RSS methods, can be assigned as the dominant contributor to these errors. 
From a different point of view, the CGM procedure contributes toward achieving higher numerical stability compared to self-consistent iterations like the IFTA method.

In contrast, the numerical and experimental values of light utilization efficiency are similar for the FOI method.
This result implies that the FOI method is tolerant to experimental disturbances concerning light utilization efficiency, which is a global property of light-spot arrays.
However, the argument does not apply to spot uniformity, which exemplifies the local properties of the spot arrays. 
The numerical results, including both the values and errors of spot uniformity, reveal distinct differences between the FOI and RSS methods, as outlined in Table~\ref{tab:var_eff}.
Thus, spot uniformity is affected by the selection of the cost function.

In practical optical systems, multiple factors such as lens aberrations, misalignment, vignetting, the uniformity and phase flatness of incident light, and phase-modulation characteristics of a SLM device interact in complicated ways to degrade the quality of the reproduced images.
Since it is almost impossible to evaluate all these factors via numerical simulation, we restricted our attention to effects of aberrations on the reproduced images. 
Numerical results in Table~\ref{tab:var_eff} show that the aberrations certainly affect significanlty the spot uniformity but do slightly the light utilization efficiency. 
It is noted again that the numerical evaluation of the aberrations is performed for informational purposes and not the subject of quantitative comparison.
Nevertheless, it is meaningful to highlight the effects of a tiny phase distortion expected in practical optical systems.

The remaining question on the results of in Table~\ref{tab:var_eff} is why there is a significant discrepancy between numerical and experimental values of the light utilization efficiency for the RSS method.
One possible qualitative explanation is as follows: In FOI-designed holograms, where diffusive components dominate, effects of extrinsic disturbance in experimental systems can be buried in the scattered elements that contribute little to holographic output images.
As a result, the FOI method presents superficial tolelance to extrinsic disturbance in practical optical systems.
In contrast, in RSS-designed holograms, where majority parts of light components contribute to the output images reproduced as results of strong interference with each other, even the minor disturbance can cause significant impact on the output images. 
In relation with the above explanation, function of FOI-designed holograms is analyzed in the next subsection. 

\subsection{Physical interpretation of FOI-designed CGHs}
The subsequent question is: \textit{how does the FOI method generate super-resolution spot arrays?} A well-known example is the phase shift method \cite{levenson1982improving} used in lithography. 
Although lithography differs from holography in terms of near and far-field light propagation phenomena, 
this technology has attempted to overcome the diffraction limits by controlling the phase distribution of coherent light. 
The RSS method can be classified under this type of approach owing to its high light utilization efficiency and globally smooth CGHs, as illustrated in Fig.~\ref{fig:hologram_images}{(b)}. 

In contrast, the FOI-designed CGH shown in Fig.~\ref{fig:hologram_images}{(b)} presents a phase profile with prominent high-frequency components that cause diffusive light propagation.
This leads to the characteristics of the FOI method, namely, low light utilization efficiency. 
Therefore, FOI-designed CGHs may function as complex amplitude modulation phase holograms \cite{Kirk1971, Davis1999, Arrizon2007, Ando2009} to remove the undesired light components that prevent super-resolution and cause interference, which induces instability in the holographically reproduced images. 
Conversely, the current results indicate a limitation of phase-only modulation approaches and the necessity of removing undesired light components to achieve super-resolution spot arrays.

\subsection{Properties of FOI cost function}
Before concluding the paper, we discuss the numerical properties of the FOI cost function. In CGH design problems, an ``absolute'' solution satisfying $I_i = T_i$ is not possible within a set of physical solutions at all instances due to the nature of light propagation.
Consequently, CGHs designed using optimization-based methods generally rely on cost functions. 
Here, it is essential to focus on the behavior of the cost function rather than its superficial form. In fact, we can readily confirmed that $\sum (\tilde{I}_i - \tilde{T}_i)^2 = 2 - 2\sum \tilde{I}_i \tilde{T}_i$ holds, indicating the equivalence of the FOI and RSS cost functions under Euclidean normalization. 
Thus, one can infer that Euclidean normalization itself plays a crucial role in generating CGHs for super-resolution spot array patterns. 
Although we were unable to establish a mathematically rigorous description of the significant changes in the reproduced images due to the cost functions, it is still worthwhile to discuss this briefly.

Consider a test target pattern such that $T_i = c > 0$ only at separated $M$ points, among $N$ lattice points in the image plane ($M \leq N$).
The support for $T_i$ is denoted by $\alpha$ as follows: $\alpha = \{ \forall i | T_i \neq 0 \}$.
The target pattern satisfies $T_i = 1/M \ (i \in \alpha)$ for the total-sum normalization, whereas $\tilde{T}_i = 1/\sqrt{M} \ (i \in \alpha)$ holds for Euclidean normalization. 
Therefore, the FOI cost function becomes 
$f_\text{FOI} = -(1/\sqrt{M}) \sum_{i \in \alpha} \tilde{I}_i$.
However, the RSS cost provides:
\begin{align}
\sum_i \left( I_i - T_i \right)^2 &= 
2 - {\sum_{i \neq k}}^\prime T_i T_k - 2 \sum_{i \in \alpha} I_i T_i 
-  {\sum_{i \neq k}}^\prime I_i I_k \notag \\
&= \left(1 + \frac{1}{M} \right) - \frac{2}{M} \sum_{i \in \alpha} I_i - {\sum_{i \neq k}}^\prime I_i I_k, \label{eq:RSScomp}
\end{align}
where $\sum I_i^2 = 1 - \sum_{i \neq k}^\prime I_i I_k$ and a similar relation for $T_i$ was used for the aforementioned deformations.
Note that the sum of the output values over $\alpha$ commonly works to reduce both FOI and RSS cost values, despite the difference in normalization. 
However, an additional factor in $f_\text{RSS}$ reduces the cost value: the last term of Equation \eqref{eq:RSScomp}. This term accounts for the autocorrelation of the output pattern independent of the target, and it broadens the light spots, which is favorable for reducing $f_\text{RSS}$.

\setcounter{lastequationbeforeappendix}{\number\value{equation}}

\section{Conclusions\protect}\label{sec:conlcusions}
This study discovered that the quality of narrow holographic optical lattice patterns could be significantly improved by utilizing the FOI method
for optimization-based hologram design,
as opposed to the traditional RSS method. We have developed a super-resolution criterion specifically for arrays of light spots, which is based on the spacing of the spots and the VP ratio of the incoherently superimposed Airy disks. 
The multispot patterns generated in the present experiments satisfied this criterion, representing a key advancement for subsequent cold atom experiments employing holographic techniques \cite{tomita2024, IMS2}. The potential applications of this method are not limited to cold atoms; it also holds promise for laser processing \cite{sakakura2010, Sakakura2011} and bioimaging \cite{nikolenko2008}.
Our method enables flexible and high-precision spot alignment without modifying the optical components used in the system, making it a versatile technique for a wide range of applications. When combined with a high-resolution optical system, our approach has the potential to generate extremely fine patterns.
In future, we intend to establish a comprehensive understanding of the relationship between other hologram design methods, such as IFTA and its derivatives, and optimization-based methods. This exploration is expected to evolve, thereby expanding the potential application areas of the proposed method.

\begin{acknowledgments}
  The authors are grateful to H.~Toyoda, T.~Inoue, and Y.~Ohtake of HPK for their encouragement throughout this work, and to T.~Watanabe for kind supports on fabrication of optics. 
  We also appreciate discussion and kind support provided by Prof. K.~Ohmori of IMS as the leader of ultrafast cold-atom quantum simulator projects. 
  Part of this study was performed under support of JSPS Grant-in-Aid for Specially Promoted Research (grant no. 16H06289). 
  T.T. and S.d.L. also acknowledge partial support by JSPS Grant-in-Aid for Research Activity Start-up (grant nos. 19K23431 and 19K23429, respectively) as well as by MEXT Quantum Leap Flagship Program (MEXT Q-LEAP) JPMXS0118069021 and JST Moonshot R\&D Program Grant Number JPMJMS2269.
\end{acknowledgments}

\appendix

\section{Incident light conditions and reproduced image quality}\label{sec:LightConditions}
In this section, we present the results of numerical simulations on how amplitude distribution of incident light affects
the quality of holographically reproduced images.
Although holograms were designed by assuming a top-hat beam as incident light, i.e., uniform illumination over a circular region around the center
of hologram plane, it is difficult in practice to generate precise top-hat beams.
For this reason, a truncated Gaussian beam is often adopted in practical experiments as a pseudo top-hat beams.
Therefore, we numerically simulated the spot uniformity, light utilization efficiencies and VP ratio when the holograms are designed with a top-hat beam
as incident light but reproduced by using a truncated Gaussian beam.
The simulation conditions not mentioned below are set to be the same as those in Sec.~\ref{sec:experiments}.

We evaluated these properties from hologram images, which are reproduced numerically under illumination of truncated Gaussian beams, 
for 20 FOI-holograms designed in Sec.~\ref{sec:experiments}. 
Illumination profiles of the truncated Gaussian beams were determied by extracting the central area of 4.5~mm radius (corresponding to 360~px
on the active area of the SLM) of Gaussian beams with $1/e^2$ radius of 5, 10, and 20~mm. 
The results are summarized in Table \ref{tab:TophatGauss} together with the values obtained from numerical results reproduced under illumiation of an ideal top-hat beam for reference.
In the case of 20~mm radius, which is close to the experimental conditions, there is no significant change in any of the properties, which suggests that it is reasonable to approximate the incident beam as a top-hat beam. On the other hand, if incident light is narrowed down to a radius of 5~mm, the quality degradation of the reproduced image becomes notable.

\begin{table}[htbp]
  \caption{Spot uniformity, light utilization efficiency, and VP ratio of light-spot arrays that are numerically reproduced
from FOI-designed CGHs illuminated by top-hat beam and truncated Gaussian beams of various radii.}
  \label{tab:TophatGauss}
\centering
\begin{tabular}{@{}lccc@{}}
\toprule
  $1/e^2$ radius & Spot uniformity & Util. eff. & VP ratio \\
\midrule
  (Top-hat) & $1.21(4) \times 10^{-2}$ & 0.120(3) &  $1.00(5) \times 10^{-2}$ \\
  20 mm & $1.41(4) \times 10^{-2}$ & 0.116(3) & $1.00(4) \times 10^{-2}$ \\
  10 mm & $2.6(1) \times 10^{-2}$ & 0.108(3) & $1.10(4) \times 10^{-2}$ \\
  5 mm & $9.9(3) \times 10^{-2}$ & 0.076(2) & $1.67(5) \times 10^{-2}$ \\
\bottomrule
\end{tabular}
\end{table}

\section{Applicability of our method to non-square lattice patterns}\label{sec:NonSquareLattice}
\begin{figure}[htbp]
  \includegraphics[width=\columnwidth]{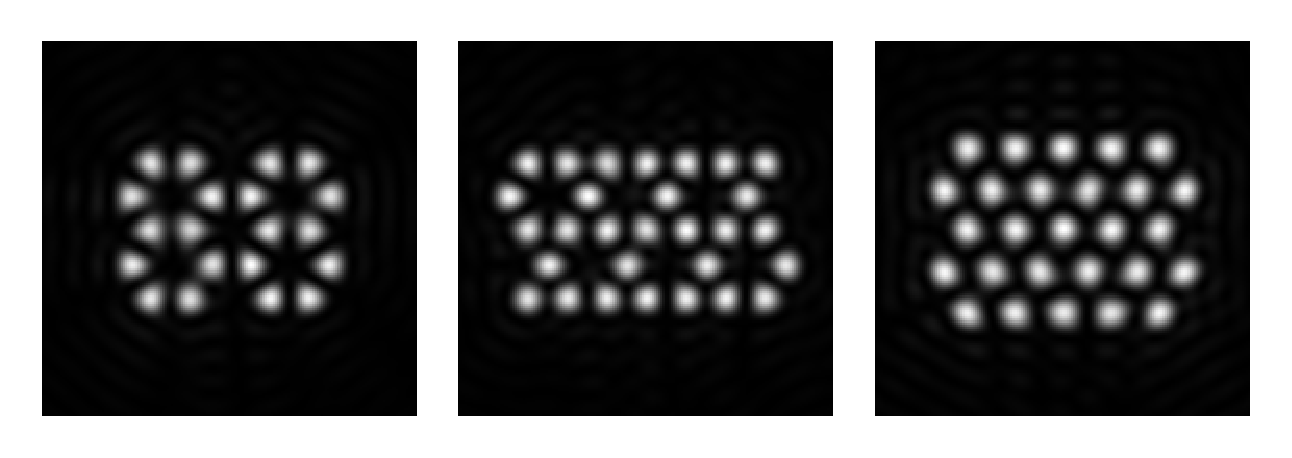}
\makeatletter\long\def\@ifdim#1#2#3{#2}\makeatother
\caption{Reproduced images of FOI-holograms with hexagonal, kagome, and triangular lattices as target patterns. The spot intervals are 1.9, 1.9, and 2.3 pixels, respectively.}\label{fig:Appx1} 
\end{figure}

For the reasons explained in Sec.~\ref{sec:criterion}, the experiments in this study were limited to square lattice patterns as the target pattern. 
However, to fully leverage the advantages of holographic techniques, it is worth discussing the applicability to other lattice patterns.
Our super-resolution criterion for multispot patterns is defined based on the average VP ratio between adjacent spots and is therefore applicable to non-square lattice patterns. 
Figure~\ref{fig:Appx1} shows numerically reproduced images of hexagonal, kagome, and triangular lattices generated by the FOI method, with spot spacings of 1.9, 1.9, and 2.3~pixels on the image plane, respectively.
Compared to square lattices, these alternative lattice patterns tend to allow for greater spacing between spots while still maintaining clear separation.
In the diffraction-limited region, multispot patterns behave not as collections of independent spots
but rather as spatially distributed patterns like higher-order transversal light modes
sometimes exhibiting non-circular spot shapes depending on the geometry of each lattice.
Although investigating the reasons behind these observations is beyond the scope of this study, one plausible hypothesis is that the square lattice pattern is relatively close to physically realizable Hermite-Gaussian modes, which contributes to its stability.

\providecommand{\noopsort}[1]{}\providecommand{\singleletter}[1]{#1}%

\clearpage

\section*{Supplemenatary Material}
\begin{center}
\begin{figure*}[t]
\includegraphics{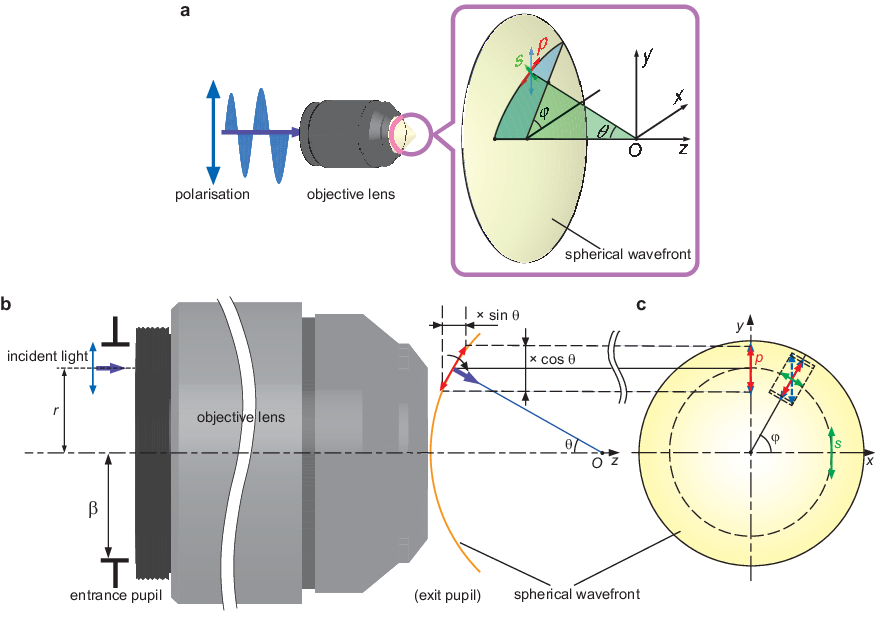} 
\caption{\label{Fig1s}
Vector nature of light amplitude in focusing.
(a)~Focusing of linearly polarized light through an objective lens. 
Vector amplitude of incident flat-wavefront light is mapped to that on spherical wavefront 
at the exit pupil of the objective lens. A position on the spherical wavefront is identified 
by the cone and directional angles, $\theta$ and $\varphi$. respectively.
(b)~Cross-sectional view of (a) on a $z-\varphi$ plane.
$\varphi$-directional component of the vector amplitude of incident light (blue arrows) is 
mapped to that (red arrows) on spherical wavefront by the objective lens. 
(c)~Front view of (a). 
The objective lens tilts the $p$-component (radial component: red arrows) of a vector light 
amplitude but does nothing to the $s$-component (azimuthal component: green arrows).
This causes inhomogeneous change in the light polarization, because the decomposition into 
the $p$- and $s$-components depends on $\varphi$ for linearly-polarized incident light.
}
\end{figure*}
\end{center}

\vspace{-4mm}
\leavevmode\\
\noindent\textbf{Vector Debye formula for focal pattern under high-NA focusing conditions.} \\
To accurately calculate the diffraction integral, the vector nature of light must be considered 
to obtain focal patterns that are deemed correct within wave optics [36, 37].
The vector nature of light in the focusing phenomenon is schematically illustrated in Fig.~\ref{Fig1s}. 
An objective lens modifies the incident light by attaching a spherical wavefront, thereby altering 
the propagation direction of light towards the center of the spherical wavefront, known as the 
geometric focus [Fig.~\ref{Fig1s}(a)]. 
As the propagation direction changes, the vector amplitude of the incident light is reduced, 
inducing a polarization component in the direction of light propagation [Fig.~\ref{Fig1s}(b)]. 
This results in an inhomogeneous modulation of the vector light amplitude due to the edge 
birefringence of the lens [Fig.~\ref{Fig1s}(c)]\cite{Inoue1952}. 
These characteristics are specific to focusing by a high-NA objective lens, and this phenomenon 
is referred to as depolarization.

The polarization vectors of light at the entrance pupil and image space of an objective lens 
are expressed in Cartesian coordinates, where the $z$-axis and origin correspond to 
the optical axis and geometrical focus, respectively [Fig.~\ref{Fig1s}(a)]. 
Although the coordinates of the entrance pupil and those in the image space are distinct, 
polarization vectors can be described in a common Cartesian coordinate system. 
In general, various coordinate systems are introduced to describe the positions on 
the entrance pupil and in the image space. 
The former is expressed in polar coordinates, $(r, \varphi)$, whereas the latter uses cylindrical 
coordinates $(\rho, \phi, z)$. 
The cylindrical and Cartesian coordinates in the image space are compatible through a coordinate 
transformation. 
Note that the azimuth ($ = \varphi$) at the entrance pupil represents the directional angle 
of the spherical wavefront at the exit pupil [Fig.~\ref{Fig1s}(a)] because an ideal objective lens 
does not alter the azimuthal direction. 
The vector Debye formula defines an integral transformation of the vector field of light amplitude 
from $(r, \varphi)$ to $(\rho, \phi, z)$, as determined by the action of a high-NA objective lens. 

\onecolumngrid
\twocolumngrid

In the context of this study, the incident light is assumed to be monochromatic, with a wavelength 
$\lambda \ (= 2 \pi / k)$. 
We assumed that the incident light has a spatially homogeneous polarization and electric field 
amplitude distribution of the form $e^{-i l \varphi} E_\text{in}(r)$. 
The focal patterns of various light beams can be derived, in principle, from a combination of 
these distributions.
The polarization vector of the incident light is expressed as a unit vector $\hat{\bm{e}}_\text{in}$, 
where the components are a pair of complex numbers, $\alpha$ and $\beta$ ($|\alpha|^2 + |\beta|^2 = 1$).  
Specifically, $\hat{\bm{e}}_\text{in} = (\alpha, \beta) = \alpha \hat{\bm{x}} + \beta \hat{\bm{y}}$ ,
where $\hat{\bm{x}}$ and $\hat{\bm{y}}$ represent unit vectors in the $x$ and $y$ directions, respectively. 
In particular, $(\alpha, \beta) = (1/\sqrt{2}, \mp i/\sqrt{2})$ describes the left or right circular polarization.

\counterwithout{equation}{section}
\setcounter{equation}{\number\value{lastequationbeforeappendix}}

When the incident light is expressed by the vector amplitude: 
$e^{-i l \varphi} E_\text{in}(r) \hat{\bm{e}}_\text{in}$ at the entrance pupil of the objective lens, 
electric field amplitude $\bm{E}$ at the observation point $(\rho, \phi, z)$ in the image space is expressed 
as follows \cite{PhysRevLett.99.073901, PhysRevA.79.033830}: 
\begin{widetext}
\begin{align}
\bm{E}(\rho, \phi, z) = \pi i^l k^2 \!\int_0^{\gamma_\text{max}} \! \! & 
\sin\theta d\theta \, {\cal E}(\theta) 
\exp\left(ikz\cos\theta\right) e^{-il\phi} \notag \\
&\times \left\{ \vphantom{\frac{1 + \cos\theta}{2}} (1 + \cos\theta) 
 J_l(k\rho\sin\theta) \hat{\bm{e}}_\text{in} \right. \notag \\
&\quad\quad -\sqrt{2} i \sin\theta 
\left[\langle \hat{\bm{e}}_\text{+}, \hat{\bm{e}}_\text{in} \rangle 
   J_{l+1}(k\rho\sin\theta) e^{-i\phi} 
- \langle \hat{\bm{e}}_\text{-}, \hat{\bm{e}}_\text{in} \rangle 
J_{l-1}(k\rho\sin\theta) e^{i\phi} \right] \hat{\bm{e}}_z \notag \\
& \left. \quad\quad + (1 - \cos\theta) 
\left[ \langle \hat{\bm{e}}_\text{+}, \hat{\bm{e}}_\text{in} \rangle 
 J_{l+2}(k\rho \sin\theta) e^{-2i\phi} \hat{\bm{e}}_\text{-} 
%
+ \langle \hat{\bm{e}}_\text{-}, \hat{\bm{e}}_\text{in} \rangle 
J_{l-2}(k\rho\sin\theta) e^{2i\phi} \hat{\bm{e}}_\text{+}
 \right] \vphantom{\frac{1-\cos\theta}{4}} \right\}, 
\label{eq:DebyeOV}
\end{align}
\end{widetext}
where ${\cal E}(\theta)$ denotes the $\theta$ distribution of the incident-light amplitude\cite{Barnett1994} 
and $\gamma_\text{max}$ denotes the maximum focusing angle [$ = \sin^{-1} (\text{NA})$]. 
Moreover, the inner product of the complex vectors is expressed as 
$\langle \bm{a}, \bm{b} \rangle$ ($= \sum_ia_i^\ast b_i$) in Eq.~\eqref{eq:DebyeOV}.

The relationship between $E_\text{in}(r)$ and ${\cal E}(\theta)$ is determined by 
the optical design of the focusing lens,
For example, 
$E_\text{in}(\beta \sin \theta / \sin \gamma_\text{max}) = {\cal E}(\theta)$ holds for 
objective lenses that satisfy the sine condition (here, $\beta$ indicates the radius 
of the entrance pupil). 
Based on equation~\eqref{eq:DebyeOV}, we can confirm that the circular-polarization flat-wavefront 
light (i.e., circularly-polarized light of $l = 0$) presents a focal pattern identical 
to the mean of those of linear-polarization (e.g., $x$- and $y$-polarizations) flat-wavefront light.

\begin{SCfigure*} 
\includegraphics{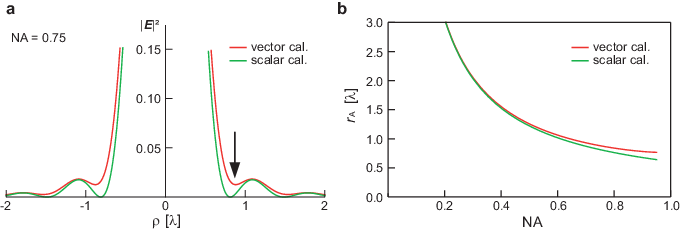} 
\caption{\label{Fig2s}
Comparison between scalar and vector calculations.
(a)~Cross-sectional profiles of focal patterns calculated by scalar
diffraction formula (green line) and vector Debye integral (red line). 
Both of the profiles were calculated under the condition of $\text{NA} = 0.75$.
(b)~Airy-disk radii as functions of NA. Results of scalar diffraction formula and 
vector Debye integral are plotted in green and red lines, respectively. 
The scale of the length is selected as the wavelength of light in both (a) and (b).
}
\end{SCfigure*}
\leavevmode\\
\noindent\textbf{Airy-disk radius as function of NA.} \\
Since the super-resolution criterion proposed in the main text is established by selecting
the radius of an Airy disk as a unit of the length, it is necessary to know the actual value 
of the Airy-disk radius for determining whether an observed light-spot array exceeds the 
super-resolution criterion or not under a particular optical condition. 
From an opposite point of view, it is possible to apply the proposed super-resolution criterion 
to another optical condition by replacing the Airy-disk radius with that derived properly 
according to the condition under consideration.

An Airy disk was originally defined as the diffraction pattern of a circular aperture in 
the far-field region of light propagation and is derived from the scalar diffraction formula, 
which is proportional to $[J_1(k \rho \text{NA})/ (k \rho \text{NA})]^2$ on the focal plane. 
The distance from the center of the image to the innermost dark ring is often referred to 
as the Airy disk radius, which is related to the conventional super-resolution criterion.

Equation~\eqref{eq:DebyeOV} reproduces a focal pattern that is similar to that of 
a conventional Airy disk. 
However, a finite value was observed in the innermost dark ring [Fig.~\ref{Fig2s}(a)]. 
Therefore, the Airy-disk radius is redefined as the distance from the optical axis 
to the innermost local minimum in the high-NA region. 
The redefined Airy-disk radius is marginally larger than the conventional one, 
which is given by $0.61/\text{NA}$ in units of wavelength. 
The conventional and high-NA Airy-disk radii are depicted as functions of NA in Fig.~\ref{Fig2s}(b) 
to illustrate the difference between them. 
This difference is almost negligible for smaller NA values but becomes notable for $\text{NA} \gtrsim 0.7$.

\leavevmode\\
\noindent\textbf{Super-resolution criteria in high-NA regions.} \\
\begin{figure}
\begin{center}
\includegraphics{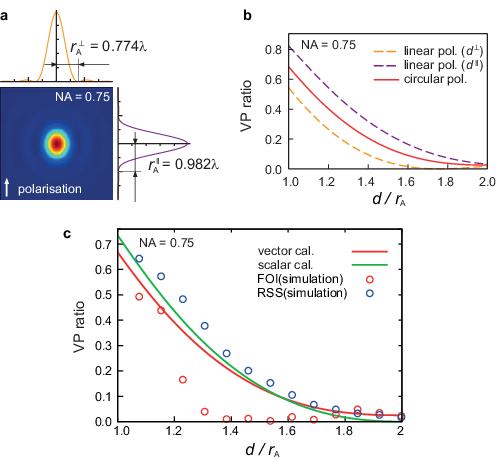} 
\end{center}
\caption{\label{Fig3s}
Effects of polarization under condition of NA = 0.75.
(a)~Focal pattern and corresponding cross-sectional profiles of linear-polarization light.
(b)~Super-resolution criteria for circular-polarization (solid red line) and 
$y$ linear-polarization (dashed lines) light. 
In the case of the linear-polarization light, the super-resolution criteria acts differently 
according to the light-polarization direction (dashed purple line) and
the perpendicular direction (dashed orange line).
(c)~Comparison of numerical spot-array patterns with the super-resolution criteria. 
Open circles present spot spacing and VP ratio of the spot-array patterns 
obtained by applying the numerical hologram reproduction procedure (FFT) to holographic 
phase patterns designed by FOI (red open circles) and RSS (blue open circles) methods.
Red and green solid curves display the super-resolution criteria calculated by the
vector and scalar diffraction formulae, respectively. 
}
\end{figure}
Linear-polarization flat-wavefront light produces a focal pattern as shown in Fig.~\ref{Fig3s}(a)
under high-NA focusing conditions. Consequently, the Airy-disk radius has different values 
in the horizontal and vertical directions. 
Therefore, different length scales must be introduced to the horizontal and vertical spacings 
in a light spot array of linear-polarization light for estimation of super-resolution according 
to the proposed criteria in the main text.

Equation~\eqref{eq:DebyeOV} indicates that a focal pattern with circular-polarization flat-wavefront 
light is determined as the sum of the horizontal and vertical components. 
This also holds true, not strictly but effectively, up to the leading three digits of the Airy-disk 
radius. 
According to this observation, we can expect that the relation between super-resolution criteria 
calculated for linearly and circularly polarized light is useful for evaluating super-resolution.
In fact, Fig.~\ref{Fig3s}(b) reveals that the super-resolution criterion for circularly polarized 
light can be approximated as the mean of those for the horizontal and vertical linear-polarization 
lights, where the Airy-disk radius of the circularly polarized light is used commonly as the scale 
of length.
Thus, the super-resolution criterion for circularly polarized light is expected to be applicable to 
the evaluation of the mean spot spacing in a square light-spot array of linearly polarized light.

To evaluate the super-resolution of a given light spot array, it is reasonable to calculate a criterion 
with an appropriate selection of the light propagation conditions under which the spot-array pattern 
was generated. 
As hologram design is performed by applying the FFT, specifically through Fraunh\"{o}fer diffraction 
calculation, the super-resolution criterion for numerically generated light-spot patterns can be 
assessed using the scalar diffraction formula. 
On the other hand, the super-resolution for an experimentally observed pattern cannot be determined 
from the criterion based on the scalar diffraction formula, because the approximate condition of 
the formula is not satisfied at a NA = 0.75.
As discussed in the main text, the VP ratio of the experimentally observed patterns exceeds 
the super-resolution criterion based on the vector Debye formula, indicating that super-resolution 
patterns were obtained using the FOI-cost method.

Nevertheless, the super-resolution criterion based on the scalar diffraction formula is expected 
to provide insights into estimating the potential effectiveness of the method. 
Figure~\ref{Fig3s}(c) suggests that the FOI-cost method can potentially produce super-resolution 
light-spot arrays based on the super-resolution criterion (solid green line) calculated using 
the scalar diffraction formula.


\end{document}